\documentclass[journal=jctc,manuscript=article]{achemso}
\usepackage{bm}
\usepackage{caption}
\usepackage{subcaption}
\usepackage{graphicx}
\usepackage{float}
\usepackage{xcolor}
\usepackage{color}
\usepackage[hidelinks,breaklinks,colorlinks=true,linkcolor=blue,citecolor=blue,urlcolor=blue]{hyperref}
\usepackage[version=3]{mhchem} 
\SectionNumbersOn

\author{Nikhil R. Agrawal}
\affiliation[University of California, Berkeley]
{Department of Chemical and Biomolecular Engineering, University of California, Berkeley, CA 94720-1462, USA}
\author{Rui Wang}
\email{ruiwang325@berkeley.edu}
\affiliation[University of California, Berkeley]
{Department of Chemical and Biomolecular Engineering, University of California, Berkeley, CA 94720-1462, USA}
\alsoaffiliation[Lawrence Berkeley National Laboratory]
{Materials Sciences Division, Lawrence Berkeley National Laboratory, Berkeley, 94720, California, USA}

\title{Electrostatic correlation induced ion condensation and charge inversion in multivalent electrolytes}


\begin{document}
\begin{tocentry}
\vspace{-1.8em}
\begin{figure}[H]
\captionsetup[subfigure]{labelformat=empty}
 \includegraphics[width=1\columnwidth]{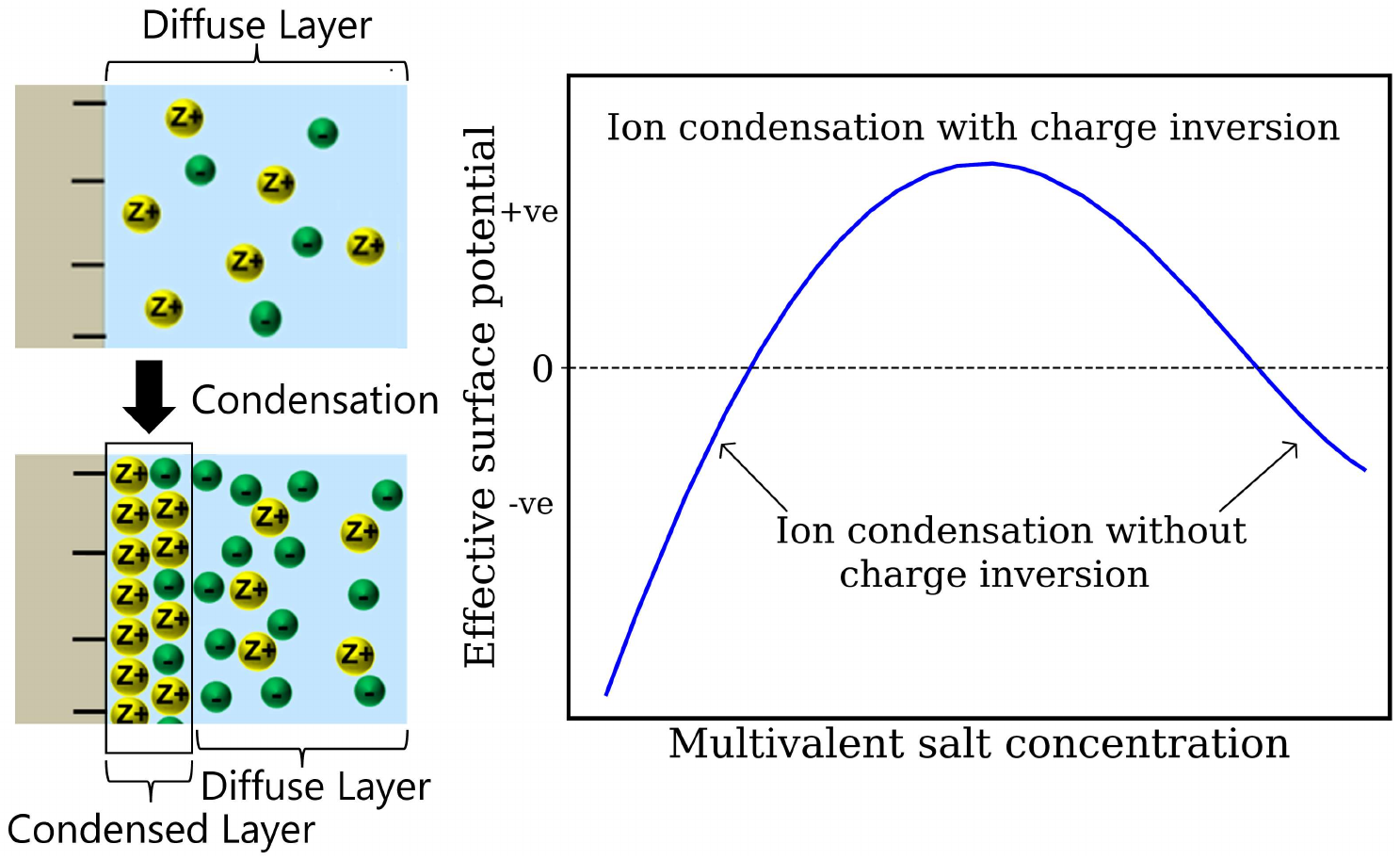}
    \caption*{}
\end{figure}
\end{tocentry}

\begin{abstract}
The study of the electrical double layer lies at the heart of colloidal and interfacial science. The standard mean-field Poisson-Boltzmann (PB) theory is incapable of modeling many phenomena originated from ion correlation. An important example is charge inversion or overcharging of electrical double layers in multivalent electrolyte solutions. Existing theories aiming to include correlations cannot capture the non-monotonic dependence of charge inversion on salt concentration because they have not systematically accounted for the inhomogeneous nature of correlations from surface to the bulk and the excluded volume effect of ions and solvent molecules.
In this work, we modify the Gaussian renormalized fluctuation theory by including the excluded volume effect to study ion condensation and charge inversion. A boundary layer approach is developed to accurately model the giant difference in ion correlations between the condensed layer near the surface and the diffuse layer outside. The theory is used to study charge inversion in multivalent electrolytes and their mixtures. We predict a surface charge induced formation of a three-dimensional condensed layer, which is necessary but not sufficient for charge inversion. The value of the effective surface potential is found to depend non-monotonically on the bulk salt concentration. Our results also show a non-monotonic reduction in charge inversion in monovalent and multivalent electrolyte mixtures. Our work is the first to qualitatively reproduce experimental and simulation observations and explains the underlying physics.
\end{abstract}

\section{Introduction}
The study of the electrical double layer is of fundamental importance to interfacial science as it affects a wealth of structural and dynamic behaviors in a variety of physicochemical, soft matter, and biophysical systems \cite{newman2021electrochemical, Levin2002ElectrostaticBiology, israelachvili2011intermolecular, Felgner1997NonviralTherapy}. In the so-called weak-coupling limit, it is generally accepted that the double-layer structure can be well described by the mean-field Poisson Boltzmann (PB) theory. However, PB theory fails to explain systems with large surface-charge density, high counter-ion valency, and high ion concentration— the strong coupling condition because it ignores the electrostatic correlation. The absence of local fluctuation effect in PB does not allow it to even qualitatively capture phenomena such as charge inversion and like charge attraction, often observed in systems with multivalent electrolytes\cite{Besteman2004DirectPhenomenon, Grosberg2002Colloquium:Systems, Kekicheff1993ChargeElectrolyte, Semenov2013ElectrophoreticSimulations, Naji2013Perspective:Beyond}. Modeling these counter-intuitive phenomena is essential to understanding DNA packing in biological organisms\cite{Gelbart2007DNAInspiredElectrostatics, Besteman2007ChargeIons}, stability of colloidal dispersions\cite{Tata2001ColloidalColloids} and dramatic change of the lubricating power in polyelectrolyte brushes\cite{Yu2018MultivalentBrushes}. \par

Charge inversion, as a result of over-accumulation of counterions in the double layer, leads to a reversal in sign of electrophoretic mobility of a charged colloidal particle\cite{Martin-Molina2008ChargeSimulations, Semenov2013ElectrophoreticSimulations} or the direction of ionic current in electrokinetic flows\cite{VanDerHeyden2006ChargeCurrents,dockalgrapheneovcharge2019}. This excess accumulation is feasible either by ion correlations or specific chemical interactions between counterions and the charged surface\cite{lyklema2006}. Over the last few decades, many studies have been conducted to study the dependence of charge inversion on these two mechanisms. Both experiments\cite{Besteman2004DirectPhenomenon} and simulations\cite{kubickovaovhargesim2012} have shown that in multivalent salt solutions, the dominant driving mechanism is the electrostatic correlation. Kub\'{\i}\ifmmode \check{c}\else \v{c}\fi{}kov\'a et al.\cite{kubickovaovhargesim2012} performed electrophoretic mobility measurements for charged colloidal particles and observed a reversal in sign of mobility upon increasing multivalent salt concentration. Using only coulombic pair potentials and hard sphere interactions for Monte Carlo simulations, they obtained good agreement with experimental results, indicating the electrostatic origin of charge inversion in multivalent salts. In addition, their simulations did not show any reversal in sign of mobility for particles dissolved in Na$^+$ and K$^+$, proving that monovalent counterions cannot overcharge the double layer in the absence of specific chemical interactions. With a smartly designed atomic force microscope (AFM) setup, Besteman et al.\cite{Besteman2004DirectPhenomenon} found that the critical multivalent salt concentration at which charge inversion occurs is only dependent on the valency of the counterions. They also conducted streaming current measurements inside nanochannels and found that the direction of the ionic current gets reversed beyond a critical concentration of multivalent counterions\cite{VanDerHeyden2006ChargeCurrents}. The magnitude of the inverted current is enhanced as the salt concentration increases, reaches a maximum, and eventually gets suppressed at high salt concentrations. Similar non-monotonic dependence of charge inversion on the salt concentration has been observed in computer simulations on electrophoresis of polyelectrolytes\cite{Hsiao2006Salt-inducedPolyelectrolytes, Hsiao2008OverchargingStudy}. Furthermore, recent experiments show that although the initial addition of multivalent ions leads to aggregation of like-charged colloids, the aggregates redissolve on continued addition of salts \cite{Asor2017CrystallizationNanoparticles, Kumar2017InteractionsIons}. This phenomenon of precipitation and resolubilization on adding multivalent salts has also been seen in polymeric systems and is commonly known as ``reentrant condensation"\cite{Saminathan1999IonicDNA, Nguyen2000ReentrantCounterions}. \par  

Despite several theoretical attempts to model ion correlations, many of the aforementioned experimental observations remain unexplained. Inspired by one component plasma physics, Perel and Shklovskii assumed the presence of a two-dimensional condensed layer of counterions on the charged surface existing in the form of a Wigner crystal (WC) lattice. The WC lattice at the surface was connected to the outside diffused layer described by mean-field PB\cite{Perel1999ScreeningInversion,rouzinabloomfieldscl1996}. This strong-coupling (SC) theory was applied to many systems consisting of charged polymers, membranes, and colloids\cite{Nguyen2000ReentrantCounterions,Shklovskii1999WignerPolyelectrolytes}. Moreira and Netz\cite{MoreiraStrong-couplingDistributions} made corrections to the SC theory by performing a perturbative expansion and compared the theoretical results with Monte-Carlo simulations. The WC picture is exact in the strong coupling limit; however, it fails to capture the onset of the ion condensation, hence missing the transition from the normal double layer to the overcharged double layer that occurs in the intermediate coupling regime. In addition, because the ion correlation and the excluded volume effect in the diffuse layer have not been included, the SC theory fails to capture the following key experimental observations. It predicts that the charge inversion will be enhanced by increasing the concentration of multivalent counterions, which cannot explain the non-monotonic dependence of charge inversion on salt concentration observed in experiments and simulations \cite{VanDerHeyden2006ChargeCurrents, Hsiao2006Salt-inducedPolyelectrolytes, Martin-Molina2002ElectrophoreticEffect}.
Furthermore, the prediction of giant charge inversion in the mixture of multivalent and monovalent electrolytes by SC theory contradicts experimental observations\cite{Nguyen2000MacroionsCharge}. Experimental and simulation studies have shown that the strength of charge inversion gets reduced as monovalent salts are added\cite{VanDerHeyden2006ChargeCurrents, Lenz2008SimulationOvercharging, Quesada-Perez2005SimulationIons}. The non-monotonic dependence of effective surface charge on monovalent salt concentration observed in the streaming current experiment also goes against the prediction of the SC theory.  \par

To describe the ion condensation on charged plates, Lau et al.\cite{Lau2002ChargeCondensation, Lau2008FluctuationSolution, Lau2002CounterionAttraction} used a point-charge model for the ions and developed a field-theory formulation based on a one-loop expansion around the saddle point value of electrostatic potential. Bazant and coworkers\cite{Bazant2011DoubleCrowding, Storey2012EffectsPhenomena, dsouzabazantcrowding} constructed a phenomenological free energy expression by writing the correlation contribution in terms of gradients of electrostatic potential and an associated correlation length parameter. Recently, Gupta et al.\cite{guptastoneovercharging} added a screening potential to the Boltzmann factor to account for ion correlations. This screening potential was also expressed in terms of the electrostatic potential gradient but without any phenomenological parameter. While all these models could predict the excess accumulation of counterions on the surface and the subsequent charge inversion, similar to the SC theory, they cannot capture the non-monotonic dependence of the charge inversion on multivalent ion concentration. Lau's point charge model overestimates ion correlations and does not include the excluded volume effects of ions and solvent molecules, thus permitting unlimited accumulation of counterions at the surface and a monotonic growth in charge inversion with salt concentration. For the models of Bazant et al.\cite{Bazant2011DoubleCrowding, Storey2012EffectsPhenomena, dsouzabazantcrowding}, and Gupta et al.\cite{guptastoneovercharging}, although excluded volume effects have been included, the gradient of electrostatic potential vanishes in bulk thus were unable to capture any correlation contribution to the bulk free energy. This incorrect description of the bulk leads to an inaccurate description of the interface, the two being in equilibrium.\par

Integral equation-based theories (IET) are another class of methods that have been widely used to address the problem of charge inversion\cite{Martin-Molina2003LookingElectrolytes, jimenezcassou2004, Martin-Molina2008ChargeSimulations,jimenezcassou2008,teranscrivenpart1,teranscrivenpart2,waismanocharge1972}. To our knowledge, no existing work based on IET has successfully described the decrease of inverted effective surface charge at high salt concentrations. While these models did account for the hard-sphere interaction between two ions as well as ions and the charged surface, their implicit treatment of solvent ignores its excluded volume, again overestimating the accumulation of counterions near the charged surface. Furthermore, from a practical standpoint, integral equations are known to be computationally expensive, with numerical stability issues often encountered at high valencies and high salt concentrations. \par

Efforts have also been made to include hydration effects of water molecules\cite{mashayakaluru2018} and ion complexes\cite{avniandelmanoss2020,jimenezcassoupairing2003,santoslevinpair2010} into the free energy. However, to our knowledge, all the existing theories fail to fully describe charge inversion because of two crucial factors that have not been systematically accounted for: the inhomogeneous nature of ion correlation from surface to the bulk and the excluded volume of both ions and solvent molecules. The inhomogeneity in ion correlation is an outcome of spatially varying ion distribution in the double layer and is particularly significant under the conditions of ion condensation. The correlation in the vicinity of the charged surface is much stronger and highly anisotropic than that in the bulk solution, although the chemical potential of ions remains the same. The existing theories cannot 
correctly model the bulk and interface simultaneously without capturing this spatially varying correlation effect. This deficiency prevents them from explaining the non-monotonic dependence of charge inversion on salt concentration and mixtures. In this work, we properly address the correlation effect by modifying the Gaussian renormalized fluctuation theory and using it to model ion condensation and charge inversion in multivalent electrolytes. The ion correlation is accounted for in the theory through the self-energy of ions. The excluded volumes of ions and solvents are included to avoid the unlimited accumulation of ions in the condensed layer. A boundary layer approach is developed to capture the giant difference in the ion correlation between the condensed layer near the surface and the diffuse layer outside. We first apply the theory to an asymmetric salt ($z:1$) solution in contact with a single charged plate to study the transition from a normal diffused double layer to a condensed double layer where charge inversion may occur. The theory is then applied to study ion condensation and charge inversion in the mixture of multivalent and monovalent electrolytes. Theoretical predictions are compared with experimental and simulation results available in the literature.

\section{Model and Theory}
\label{sec:theory}

To capture ion-ion correlations, it is necessary to go beyond the mean-field PB theory by considering the fluctuation of the electrostatic field. Using the Gaussian renormalized variational
approach, a general theory for electrolyte
solutions with dielectric inhomogeneity have been previously developed by Wang\cite{Wang2010FluctuationEnergy}. This section first modifies the general theory by including the excluded volumes of both ions and solvents in the partition function. The resulting equations are then applied to the case of electrolyte solutions in contact with a single charged plate. Finally, a boundary layer approach is introduced to account for the giant difference in the ion correlation between the vicinity of the surface and the bulk phase when ion condensation happens. This theoretical framework and numerical approach can also be applied to describe double layer structure and force between two charged plates.  

\subsection{General Theory}
Here, we derive a general theory to describe the behavior of electrolyte solutions with spatially varying ion correlations and dielectric permittivity. The system consists of a fixed charge density $\rho_\mathrm{ex}(\bf{r})$, and mobile cations and anions with valencies $q_{+}$ and $q_{-}$, respectively. The fixed charge and the mobile ions exist together in a medium with the dielectric function given by $\varepsilon ({\bf r})$.
Instead of using the point-charge model to describe the ions, we employ a more realistic description by accounting for finite charge spread by using a short-range distribution function $h_{\pm}({\bf r}-{\bf r}_i)$
for the i$^\mathrm{th}$ ion. The introduction of finite charge distribution on the ion avoids overestimating the ion correlation, which results from the point-charge model. Both ions and solvent molecules are treated spheres of finite volume in order to avoid unlimited accumulation in the condensed layer. The excluded volumes of ions and solvents are taken to be $v_{\pm}$ and $v_\mathrm{s}$, respectively, to account for the steric effect.\par

The total charge density at any position $\mathbf{r}$ in the system is
\begin{eqnarray}
e\rho({\bf r}) = e\rho_{ex}({\bf r}) +e\int d{\bf r}' \left[ q_+h_{+} ({\bf r}-{\bf r}')\hat{c}_{+}({\bf r}') -q_- h_{-} ({\bf r}-{\bf r}')\hat{c}_{-}({\bf r}')\right] 
\end{eqnarray}
where $ \hat{c}_{\pm} ({\bf r}) = \sum_{i=1}^{n_{\pm}} \delta ({\bf
r}-{\bf r}_i) $ is the particle density operator for the ions, and $e$ is the elementary charge. The electrostatic Hamiltonian, $H$, can be written as
\begin{eqnarray}
{H}= \frac{e^2}{2}\int d{\bf{r}}d{\bf{r'}}\rho({\bf{r}})C({\bf{r}},{\bf{r'}}))\rho(\bf{r'})
\end{eqnarray} 
$C(\bf{r},\bf{r'})$ is the Coulomb operator given by
\begin{eqnarray}
{-\nabla.[\epsilon(\mathbf{r})\nabla C(\mathbf{r},\mathbf{r'})]}= \delta({\mathbf{r} - \mathbf{r'}})
\label{eq:couloumb}
\end{eqnarray} 
$\epsilon({\bf{r}})=kT\varepsilon_{0}\varepsilon ({\bf r})/e^2$ is the scaled permittivity with $\varepsilon_{0}$ the vacuum permittivity. The grand canonical partition function $\Xi$ for the aforementioned Hamiltonian can be written as  
\begin{eqnarray}
\Xi=\sum_{n_{{+}}=0}^{\infty}\sum_{n_{{-}}=0}^{\infty}\sum_{n_{{s}}=0}^{\infty}\frac{{\rm e}^{\mu_+n_+}{\rm e}^{\mu_{-}n_-}{\rm e}^{\mu_{s}n_s}}{{n_+!}{n_-!}{n_s!}{v_+^{n_+}}{v_-^{n_-}}{v_s^{n_s}}}\int\prod_{i=1}^{n_+}d\bf{r}_i\prod_{j=1}^{n_-}d\bf{r}_j\prod_{k=1}^{n_s}d\bf{r}_k
\nonumber \\\times\prod_{\bf{r}}\delta\left[1-v_+\hat{c}_+(\textbf{r})-v_-\hat{c}_-(\textbf{r})-v_s\hat{c}_s(\textbf{r}))\right]{\rm {exp}}(-\beta \cal H)
\label{eq:xi}
\end{eqnarray}
where $\mu_{\pm}$ and $\mu_s$ denote the chemical potentials of the ions and solvent, respectively. The functional delta is introduced to impose the local incompressibility constraint. We then apply the Hubbard-Stratonovich transformation and identity transformation on eq \ref{eq:xi} to introduce a field variable $\phi(\bf{r})$, coupled with $\rho(\bf{r})$ and a field variable $\xi(\bf{r})$ to enforce the impressibility constraint, which yields
\begin{eqnarray}
{\Xi=\frac{1}{Z_0}\int D\phi\int D\xi \exp \left\{-L[\phi,\xi]\right\}} 
\label{eq:xi_trns}
\end{eqnarray}
The ``action" $L$ is of the following form
\begin{eqnarray}
  \ L = \int d{\bf{r}}\left[\frac{1}{2}\epsilon(\nabla\phi)^2 +i\rho_{ex}\phi-\lambda_{+}{\rm e}^{-i {\hat{h}_{+}}\phi-iv_{+}\xi}-\lambda_{-}{\rm e}^{i{\hat{h}_{-}}\phi-iv_{-}\xi} -\lambda_{s}e^{-iv_{s} \xi} - i\xi  \right] 
  \label{eq:L_var}
\end{eqnarray}
with $\lambda_{\pm}={\rm e}^{ \mu_{\pm}}/ v_{\pm}$ and $\lambda_{s}={\rm e}^{ \mu_{s}}/ v_{s}$ as the fugacity of the
ions and solvent, respectively. The short-hand notation ${\hat{h}_{\pm}}\phi$ stands for local spatial averaging of $\phi$ by the charge distribution function: ${\hat h}_{\pm} \phi= \int d{\bf r}' h_{\pm}
({\bf r}'-{\bf r}) \phi ({\bf r}')$. $Z_{0}$ in eq \ref{eq:xi_trns} is the normalizing factor expressed as
\begin{equation}
Z_0= \int D \phi \exp \left[- \frac{1}{2}  \int d {\bf r} \epsilon
(\nabla \phi)^2  \right] = \left[\det  {\bf C}\right]^{1/2}
\label{eq:zo}
\end{equation}

In this work, we focus on the fluctuation of the electrostatic field, whereas the density fluctuation is neglected. We thus take the saddle point approximation for $\xi$, setting $\eta=i\xi^*$ with $\xi^*$ denoting the saddle point value of $\xi$, which yields
\begin{eqnarray}
 v_{+}\lambda_{+}{\rm e}^{-i{\hat{h}_{+}}\phi-v_{+}\eta} + v_{-}\lambda_{-}{\rm e}^{i{\hat{h}_{-}}\phi-v_{-}\eta} +v_s\lambda_{s}e^{-v_{s} \eta} - 1 = 0
 \label{eq:eta_saddle}
\end{eqnarray}
The partition function can then be simplified to
\begin{eqnarray}
{\Xi=\frac{1}{Z_0}\int D\phi \exp \left\{-\int d{\bf{r}}\left[\frac{1}{2}\epsilon(\nabla\phi)^2 +i\rho_{ex}\phi-\lambda_{+}{\rm e}^{-i{\hat{h}_{+}}\phi-v_{+}\eta}-\lambda_{-}{\rm e}^{i{\hat{h}_{-}}\phi-v_{-}\eta} -\lambda_{s}e^{-v_{s} \eta} - \eta  \right] \right\}} 
\label{eq:xi_var}
\end{eqnarray}
The action $L$ is now a functional of $\phi$ only. To capture the fluctuation of the electrostatic potential, a variational calculation of eq \ref{eq:xi_var} is carried out using the Gibbs-Feynman-Bogoliubov bound for the grand free energy, $W$. Using this bound, $W$ can be evaluated by
\begin{eqnarray}
W = -\ln \Xi \leq {-\ln \Xi_{ref}} + \langle L[\phi] - L_{ref}[\phi] \rangle
\label{eq:gfbbound}
\end{eqnarray}
$\langle \cdots \rangle$ represents the average taken in the reference ensemble. The reference action $L_{ref}$ is chosen to be of the Gaussian form centered around the mean potential $-i\psi$,
\begin{eqnarray}
L_{ref}=  \frac{1}{2}\int d {\bf{r}}d{\bf{r'}}[\phi({\bf{r}}) + i\psi({\bf{r}})]G^{-1}({\bf{r}},{\bf{r'}})[\phi({\bf{r'}}) + i\psi ({\bf{r'}})]
\label{eq:lref}
\end{eqnarray}
where $G^{-1}$ is the inverse of the Green's function $G$, which describes the correlation between two point charges located at ${\bf{r}}$ and ${\bf{r'}}$. It is worth mentioning that by taking the saddle-point approximation for $\phi$ instead of the variational calculation shown above, the steric PB theory derived by Borukhov et al.\cite{borukhovandelmansteric1997} is recovered. Moreover, if the functional delta in eq \ref{eq:xi} is also removed, then we get back to the standard mean-field PB theory without the ion correlation and excluded volume effect of molecules.\par

Gaussian reference action leads to analytical expressions for all the terms on the r.h.s. of eq \ref{eq:gfbbound}. We refer interested readers to the relevant literature for the detailed derivation \cite{Wang2010FluctuationEnergy,Netz2003VariationalSystems}. The lower bound of the free energy is obtained by minimizing $W$
with respect to both the mean potential $\psi$ and the Green's function $G$, which results in the following two equations:
\begin{eqnarray}
{-\nabla\cdot[\epsilon\nabla\psi({\bf{r}})]}= \rho_{ex}+ q_+{\lambda_{+}{\rm e}^{-q_{+} \psi-u_{+}-v_{+}\eta}}- q_-{\lambda_{-}{\rm e}^{q_{-} \psi-u_{-}-v_{-}\eta}}
\label{eq:psi}
\end{eqnarray}
\begin{eqnarray}
{-\nabla\cdot[\epsilon\nabla G({\bf{r}},{\bf{r'}})]} + 2I({\bf{r}})G({\bf{r}},{\bf{r'}}) = \delta({\bf{r}}-{\bf{r'}})
\label{eq:greens}
\end{eqnarray} 
where 2$I({\bf r})=\left(q_+^2 c_+({\bf r})+q_-^2 c_-({\bf r})\right)$ is
the local ionic strength. The concentration of cations and
anions is given by
\begin{equation}
c_{\pm} ({\bf r}) =\lambda_{\pm} \exp \left[ \mp q_{\pm} \psi
({\bf r}) -u_{\pm} ({\bf r}) -v_{\pm}\eta ({\bf r}) \right] 
\label{eq:conc}
\end{equation}
$u_{\pm}$ above denotes the self energy of cations and anions,
\begin{eqnarray}
u_{\pm} ({\bf r})= \frac{1}{2}   \int d {\bf r}'  d {\bf r}''
h_{\pm} ({\bf r'}-{\bf r}) G({\bf r}',{\bf r}'') h_{\pm} ({\bf
r}''-{\bf r}) 
\label{eq:selfe}
\end{eqnarray}

We note that $\mu_s$ and $\eta$ are not independent of each other in an incompressible system but are intrinsically connected by the solvent concentration, i.e. $c_s=e^{\mu_s-v_s\eta}$. Without loss of generality, $\mu_s$ can be set to 0, and by making use of the incompressibility condition on $\xi$ (eq \ref{eq:eta_saddle}), an expression for $\eta$ can be obtained as
\begin{equation}
\eta(\mathbf{r}) = -\frac{1}{v_s}{\ln[1 - v_+c_+({\bf{r}}) -v_- c_-({\bf{r}})]}
\label{eq:eta}
\end{equation}

Equations \ref{eq:psi}-\ref{eq:eta} are the key equations to perform a self-consistent calculation of the mean electrostatic potential $\psi({\bf{r}})$, the correlation function $G({\bf{r}}, {\bf{r}}')$, the self-energy of ions $u_{\pm} ({\bf r})$, the ion concentration $c_{\pm}({\bf{r}})$ and the incompressibility field $\eta$. Particularly, the Green function describes the interaction between the ion and its ionic atmosphere, which manifests itself in the self energy of ions appearing in the Boltzmann factor. The self-energy can be spatially varying, which captures the inhomogeneous nature of the ion correlation effect. Based on the variational conditions (eqs \ref{eq:psi} and \ref{eq:greens}), the equilibrium grand free energy can be calculated using the charging method\cite{Wang2015OnSurfaces}:
\begin{eqnarray}
W=  \int d{\bf{r}}\left[\frac{1}{2}\psi\left(\rho_{ex} - q_+c_+ +q_-c_- \right) -c_+-c_-+ \frac{\ln(1 - c_+v_+-c_-v_-)-(1 - c_+v_+-c_-v_-)}{v_s} \right]\nonumber
\\+  \frac{1}{2}\int d{\bf{r}}\int d{\bf{r'}}\int d{\bf{r'^{\prime}}}\int_{0}^{1}d\tau[G({\bf{r'}},{\bf{r'^{\prime}}},\tau) - G({\bf{r'}},{\bf{r'^{\prime}}})]\left[\sum_{i=+,-}q_i^2c_i h_i({\bf{r'}}-{\bf{r}})h_i({{\bf{r}}-\bf{r'^{\prime}}})\right]
\label{eq:minfe}
\end{eqnarray}
where the last term in the first line of the r.h.s of eq \ref{eq:minfe} comes from the incompressibility constraint. The second line of the r.h.s of eq \ref{eq:minfe} is the contribution from the fluctuation of the electrostatic field with $\tau$ being a ``charging" variable. Here, the theory is illustrated by using the single salt as an example; however, it can be easily generalized to solutions containing multiple salts, i.e., electrolyte mixtures.\par

\subsection{Electrolyte Solutions in Contact with Charged Plates}

We now apply the general theory to the case of electrolyte solutions in contact with charged plates. The plates are taken to be infinitely large in both the $x$ and $y$ directions; the electrostatic potential $\psi$ and ion concentration $c_{\pm}$ are therefore only a function of the coordinate $z$ perpendicular to the surface. We consider the case of a single charged plate at $z=0$. The external charge density can be written as $\rho_{ex}({\bf{r}})=\sigma \delta(z)$ with $\sigma$ the surface charge density of the plate. In the solvent region ($z>0$), eq \ref{eq:psi} becomes
\begin{eqnarray}
-\epsilon_{s}\frac{d^2 \psi (z)}{dz^2}= q_+{\lambda_{+}{\rm e}^{-q_{+} \psi(z)-u_{+}(z)-v_{+}\eta(z)}}- q_-{\lambda_{-}{\rm e}^{q_{-} \psi(z)-u_{-}(z)-v_{-}\eta(z)}}
\label{eq:psi_z}
\end{eqnarray}
with the boundary condition $(d \psi / d z)_{\mathrm{z=0}}=-\sigma / \epsilon_\mathrm{s}$ at the plate surface. $\epsilon_{s}$ is the scaled permittivity of the solvent, which is assumed to be constant in the solvent region. The contribution of the salt ions to the permittivity is not considered in this work. Also, to focus on the ion-ion correlation effect, the dielectric permittivity of the charged plate is assumed to be the same as that of the solvent, such that the long-range image charge effect does not need to be considered in the current system. \par 
To obtain the self energy $u_{\pm} (z)$, we need to solve the Green function in the neighborhood of the point $z$ as shown in eq \ref{eq:selfe}. In planar geometry, it is advantageous to work in cylindrical coordinates ($\rho,z$) by applying the Hankel transform on $G(\mathbf{r'},\mathbf{r''})$ as
\begin{eqnarray}
    {G(\rho,z',z'')} = \frac{1}{2\pi}\int_{0}^{\infty} kdk J_{0}(k\rho)\hat{G}(k,z',z'')
    \label{eq:ftran-g}
\end{eqnarray}
where $J_{0}$ is the zeroth order Bessel function. $\hat{G}(k,z',z'')$ satisfies
\begin{eqnarray}
    - \frac{\partial^2 \hat{G}(k,z',z'')}{\partial z'^2}
   + (\kappa^2(z') + k^2)\hat{G}(k,z',z'') 
        = \frac{\delta(z',z'')}{\epsilon_s} 
    \label{eq:greens-ftransfd}        
\end{eqnarray} 
$\kappa(z')=\left[2I(z')/\epsilon_{s}\right]^{1/2}$ is the inverse of the local Debye screening length. To simplify the numerical calculation of $\hat{G}$, we invoke a WKB (Wentzel-Kramers-Brillouin) like approximation as in previous works \cite{Buff1963StatisticalProperties, Xu2014Self-energy-modifiedApproaches,Wang2013EffectsForces}. Under WKB approximation, Equation \ref{eq:greens-ftransfd} is first solved for a constant $\kappa$, but in the resulting expression $\kappa$ is replaced by its local value $\kappa (z)$ that depends on the local ion concentrations $c_{\pm}(z)$. $\hat{G}(k,z',z'')$ is thus given by:
\begin{eqnarray}
    \hat{G}(k,z',z'') = \frac{e^{-p(z)|{z'}-{z''}|}}{2\epsilon_s p(z)}
    \label{eq:gwkb}
\end{eqnarray}
with $\ p(z) = (k^2 + \kappa^2(z))^{1/2}$. $G(\mathbf{r'},\mathbf{r''})$ is then obtained by inverting $\hat{G}$ as in eq \ref{eq:ftran-g}:
\begin{eqnarray}
{G(\mathbf{r'},\mathbf{r''})}  =  \frac{e^{-\kappa(\mathbf{r})|\mathbf{r'}-\mathbf{r''}|}}{4\pi\epsilon_s|\mathbf{r'}-\mathbf{r''}|}
\label{eq:gsol}
\end{eqnarray}
We note that $G(\mathbf{r'},\mathbf{r''})$ in eq \ref{eq:gsol} has the same functional form as in the Debye-H\"{u}ckel (DH) theory; however, the bulk screening constant in the  DH theory is replaced by the local value which captures the inhomogeneous correlation effect.\par

As alluded to in Sec. 2.1, a finite charge spread on the ion is necessary to avoid the overestimation of correlation contribution to the self energy. 
In principle, the functional form of charge spread $\textit{h}_{\pm}(\mathbf{r}-\mathbf{r'})$ can be arbitrary. Here, we choose $\textit{h}_{\pm}(\mathbf{r}-\mathbf{r'})$ to be Gaussian distribution for mathematical convenience,
\begin{eqnarray}
{h_{\pm}(\mathbf{r} -\mathbf{r'})}  =  \left({\frac{1}{2a_{\pm}}}\right)^{3/2}\exp\left[\frac{-\pi{}(\mathbf{r}-\mathbf{r'})^{2}}{2a^2_\pm}\right]
\label{eq:cspread}
\end{eqnarray} 
where $a_\pm$ is the Born radius of the ion. Substituting equations \ref{eq:gsol} and \ref{eq:cspread} into Equation \ref{eq:selfe} yields the following expression for the self energy of ions
\begin{eqnarray}
u_{\pm}({z}) = \frac{q^2_\pm}{8\pi\epsilon_sa_\pm}-\frac{q^2_\pm \kappa(z)}{8\pi\epsilon_s}\exp\left(\frac{{a_\pm^2\kappa(z})^2}{\pi}\right)\mathrm{erfc}\left(\frac{{a_\pm\kappa(z})}{\sqrt{\pi}}\right)
\label{eq:selfe_gauss}
\end{eqnarray}
The first term on r.h.s of eq \ref{eq:selfe_gauss} is the Born solvation energy. It is a constant for uniform dielectric constant, which can be absorbed into a reference chemical potential and has no influence on the ion distribution. With the above analytical expression of self energy in hand, the electrostatic potential $\psi(z)$ and ion distribution $c_{\pm}(z)$ can be obtained by solving the self-consistent equations iteratively. \par 

\subsection{Boundary Layer Approach}

For low surface charge density, low salt concentration, and low valency-the so-called weak coupling condition, the electrostatic potential and ion distribution can be directly solved by discretizing the solution region with a finite difference method and initializing the ion concentrations using the bulk values. However, this numerical scheme could not converge as the surface charge density or salt concentration increases in multivalent ion systems. We attribute this to a high accumulation of ions near the plate surface, such that the bulk ion concentration values as initial guesses can no longer lead to a stable solution. This numerical instability indicates the formation of a correlation-induced  condensed layer with very high ion concentration at the surface. The liquid-like condensed layer is coexistent with the diffuse (gas-like) region ultimately connected to the bulk phase. This coexistence between the liquid and gas region near the charged surface is analogous to the well-known phase coexistence in bulk ionic fluids\cite{Fisher1993CriticalityBeyond}. \par

Unlike the diffused region, inside the liquid layer, there is a large amount of net charge and hence a sharp gradient in the electrostatic potential profile. To accurately model this two-region problem, a boundary layer (BL) approach similar to the tool used in fluid dynamics is necessary. In this approach, the condensed layer is solved separately like a BL, using saturated values ($1/v_{\pm}$) as the initial guess for the ion concentration. The diffused region outside the BL is solved using bulk salt concentration as the initial guess. The inner and outer solutions are connected through appropriate boundary conditions at the interface between the two regions. The first one comes from the continuity of electrostatic potential, $\psi_\mathrm{z=d_-} = \psi_\mathrm{z=d_+}$. The other boundary condition comes from the continuity of the electrostatic displacement, i.e., $(d\psi/dz)_\mathrm{z =d_-} =  (d\psi/dz)_\mathrm{z=d_+}$.
Here, $d$ is the thickness of the boundary layer, which in our approach, is a variational parameter. The equilibrium value of $d$ can be obtained from the minimization of the free energy with respect to $d$. \par

\section{Results and Discussions}
\label{sec:Results}
Here, we first apply the theory derived in Section 2 to study the electrostatic potential and concentration profiles for a single asymmetric salt solution in contact with a negatively charged plate. And then, the effect of adding monovalent salt to a multivalent salt solution will be investigated. In the current work, we focus on the effect of inhomogeneous ion-ion correlations on the double layer structure and properties. Hence, the dielectric permittivity is set to be constant even though the effect of spatially varying dielectric permittivity can be straightforwardly included in our theory.

\subsection{Onset of ion condensation and charge inversion}

In this subsection, we focus on the effect of surface charge density and  salt concentration on the double layer properties. The electrolyte is chosen to be 2:1 ($q_+=2$ and $q_-$=1). The Born radius $a_{\pm}$ is set to be 1.6 \AA. For simplicity, the radius for the excluded volumes of ions and solvent are taken to be the same: $a_{0}=2.5$ \AA\ and thus $v_{\pm,s}=\frac{4}{3}\pi(2.5)^3$ \AA$^3$. The radius for the excluded volume of ions is chosen to be larger than the Born radius to account for the hydration shell. \par

\begin{figure}
\centering
 \includegraphics[width=0.6\columnwidth]{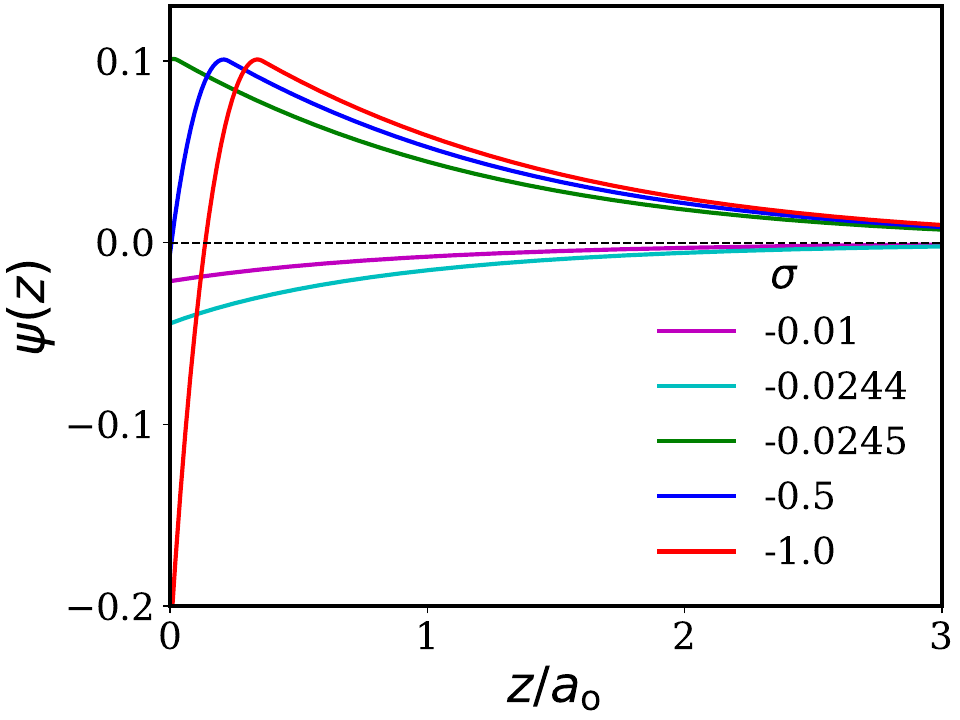}
    \caption{The electrostatic potential profiles of 2:1 electrolyte solution near a single negatively charged plate for increasing surface charge density $\sigma$ (in e/nm$^2$). The bulk salt concentration $c_{\mathrm{b}}$ is 0.2 M. The charge inversion occurs for $|\sigma| > 0.0244$ e/nm$^{2}$.}.
         \label{fig:psi_sigma02}
\end{figure}

As surface charge density increases, more counterions will be attracted to the surface region, enhancing the ionic correlation, which may lead to the instability of the normal diffuse double-layer structure. Figure \ref{fig:psi_sigma02} shows the effect of the surface charge density $\sigma$ on the electrostatic potential profile $\psi(z)$ for bulk salt concentration $c_{\mathrm{b}}=0.2$ M. For small $\sigma$ (i.e., $|\sigma| < 0.0245$ e/nm$^{2}$), $\psi$ becomes more negative as $\sigma$ increases, similar to the behavior predicted by the mean-field PB theory. In this regime, the correlation effect only introduces a quantitative correction to the PB results, as predicted by previous works using weak-coupling theories \cite{Netz2003VariationalSystems,Wang2015OnSurfaces, BuyukdagliVariationalNanopores,Attard1988ElectricalStudy}. Interestingly, as $\sigma$ further increases to a critical value of -0.0245 e/nm$^{2}$, the numerical calculation using the bulk salt concentration $c_{\mathrm{b}}$ as the initial guess cannot converge. We have to invoke the boundary layer approach as introduced in Section 2.3. A slight increase of $\sigma$ from -0.0244 e/nm$^{2}$ to -0.0245 e/nm$^{2}$ leads to a giant change in the potential profile as shown in Figure \ref{fig:psi_sigma02}. Even the sign of $\psi$ is reversed from negative to positive around this critical point: charge inversion occurs.\par 

\begin{figure}
\captionsetup[subfigure]{labelformat=empty}
    \begin{subfigure}{0.49\columnwidth}
    \includegraphics[width=\columnwidth]{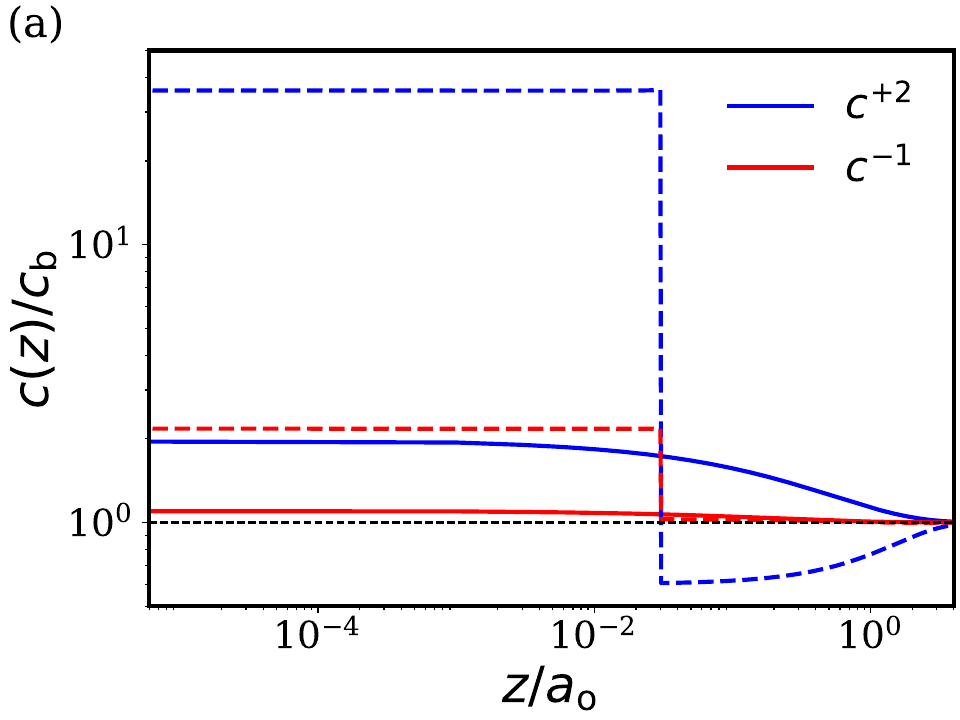}
    \caption{}
    \label{fig:conc_sigma02}
    \end{subfigure}  
    \hfill
    \begin{subfigure}{0.49\columnwidth}
    \includegraphics[width=\columnwidth]{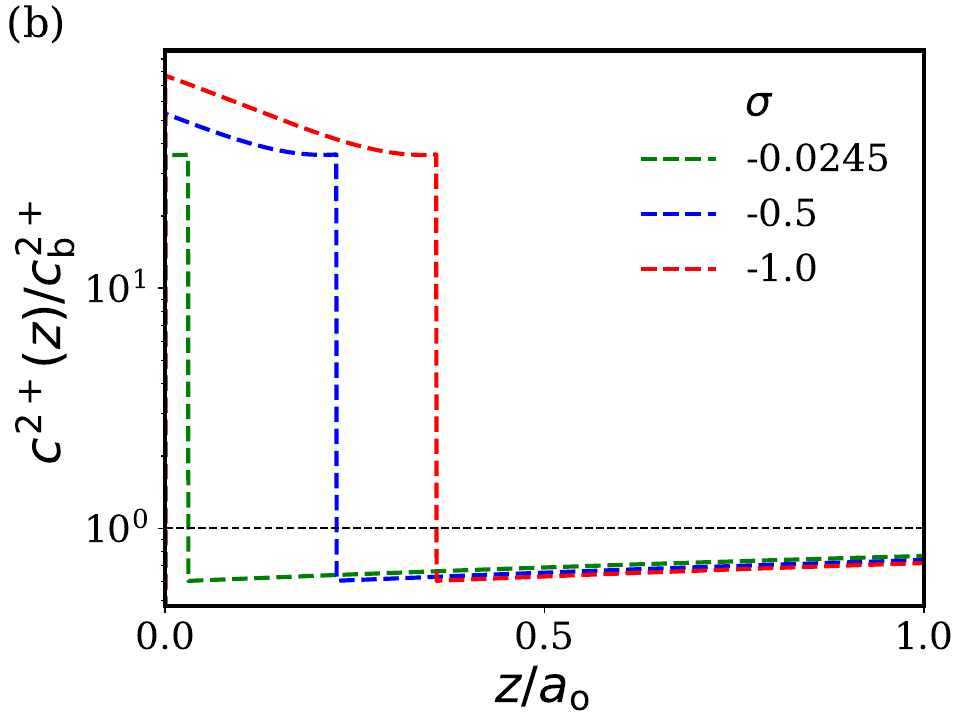}
    \caption{}
    \label{fig:conc_sigmac2}
    \end{subfigure} 
\caption{Double layer structure for ion condensation of a 2:1 electrolyte solution near a single negatively-charged plate with charge inversion, $c_{\mathrm{b}} = 0.2$ M. a) Counterion and coion distribution before ($\sigma = -0.0244$ e/nm$^2$, solid lines) and after ($\sigma = 0.0245$ e/nm$^2$, dashed lines) the ion condensation, b) Counterion distribution profiles for increasing values of $\sigma$ after the formation of condensed layer. }
\label{fig:first_panel}
\end{figure}

The charge inversion is an outcome of the ion condensation in the boundary layer region. A large number of counterions and coions are accumulated next to the charged surface to form a three-dimensional (3D) liquid-like condensed layer. The thickness of this 3D layer is of the order of ion radius. These accumulated counterions exceed the required amount to neutralize the surface charge, giving rise to charge inversion. The condensed layer structure can be more clearly seen if the ion distribution is plotted on a log scale, as shown in Figure \ref{fig:conc_sigma02}. The structure of a normal double layer before the condensation is also plotted for comparison. Compared to the case of a normal double layer, the counterion concentration $c^\mathrm{2+}$ is an order of magnitude higher in the condensed layer. Outside the condensed layer (i.e., diffuse layer region), the concentration of coion is larger than that of the counterion due to charge inversion. The coions in the region outside the condensed layer screen the effective inverted surface charge. If the condensed layer is combined with the originally negatively-charged surface as an effective surface, the inverted double layer is equivalent to a double layer near a surface with an effective positive charge. Then, the electrostatic potential at the end of the boundary layer $\psi(z=d)$ can be interpreted as the effective surface potential, $\psi_{\mathrm{eff}}$. For a stagnant condensed layer (i.e., high viscosity due to the strong inter-ion forces), 
this effective surface potential corresponds to the Zeta potential measured in electrokinetic experiments. \par

When the surface charge density further increases, $|\sigma| > 0.0245$ e/nm$^{2}$, the double layer remains overcharged. As shown in Figure \ref{fig:psi_sigma02}, the sign of $\psi_{\mathrm{eff}}$ remains positive, which is opposite to the sign of the original surface charge. There is a dramatic change in the potential within the condensed layer. Figure 2b shows that the thickness of the condensed layer grows as $\sigma$ increases due to the accumulation of more ions to the surface. The effective surface potential is found to be not sensitive to the value of $\sigma$ in this regime.

\subsection{Relationship between condensation and inversion: the effect of salt concentration}
\begin{figure}
\captionsetup[subfigure]{labelformat=empty}
    \begin{subfigure}{0.495\columnwidth}
        \includegraphics[width=\columnwidth]{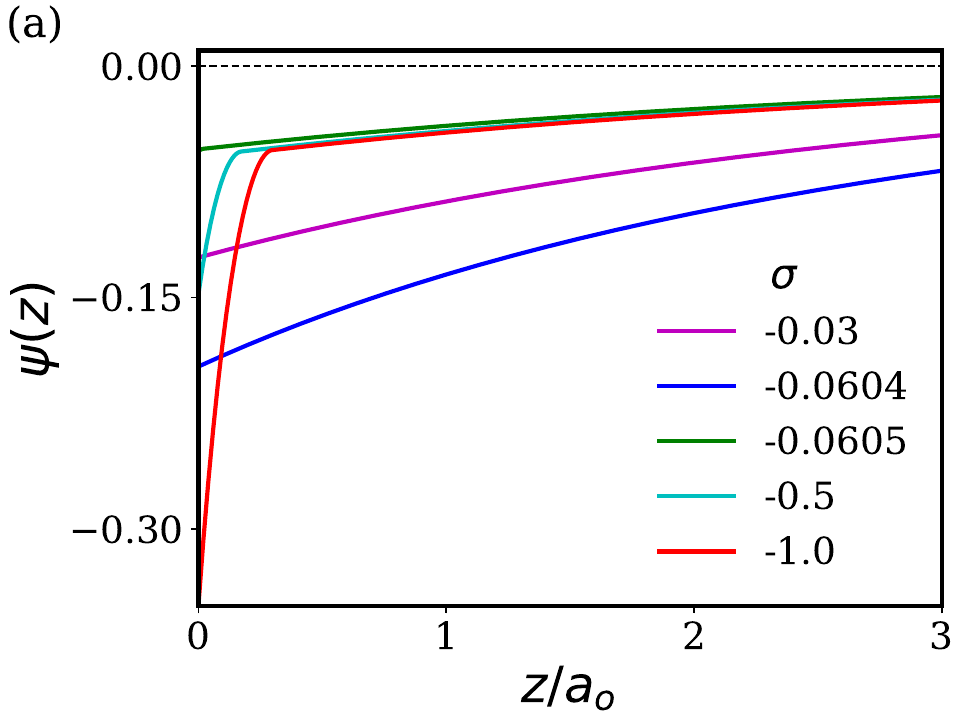}
            \caption{}
        \label{fig:psi_sigma04}
    \end{subfigure}  
       \hfill
    \begin{subfigure}{0.495\columnwidth}
        \includegraphics[width=\columnwidth]{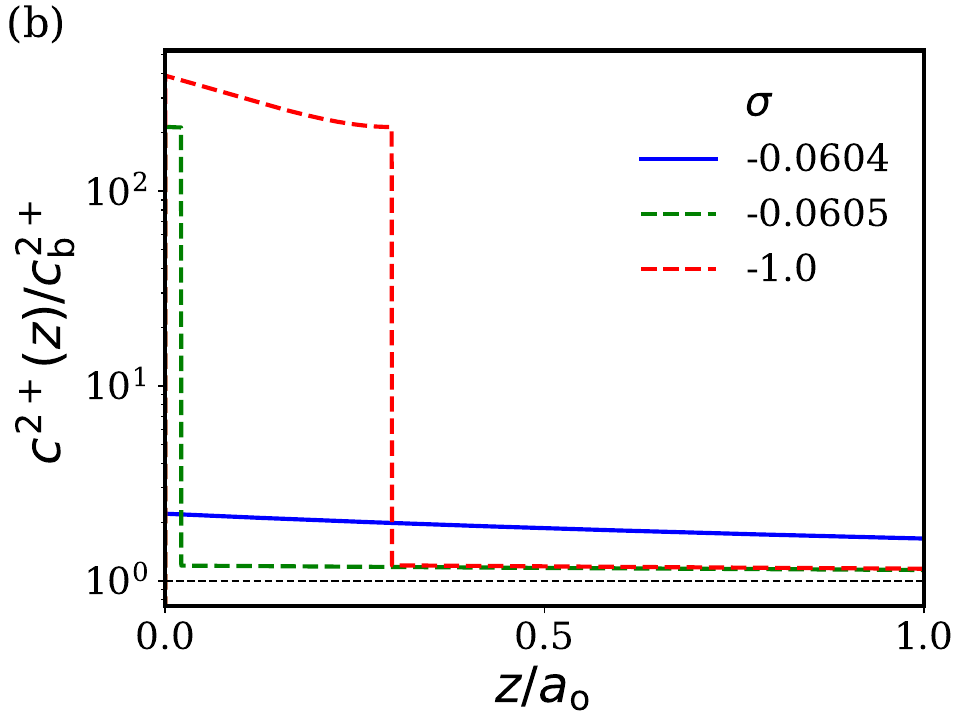}
            \caption{}
        \label{fig:conc_sigmac04}
    \end{subfigure} 
    \caption{Double layer structure for ion condensation without charge inversion. a) Electrostatic potential and b) counterion distribution for $c_{\mathrm{b}}$ = 0.04 M.}
\label{fig:panel04}
\end{figure}
In Section 3.1, the synchronic occurrence of ion condensation and charge inversion has been observed. It is natural to ask whether ion condensation necessarily leads to charge inversion. To address this question, double-layer structures are investigated at different bulk salt concentrations. For the case of a dilute solution with $c_{\mathrm{b}}=0.04$ M, condensation similar to $c_{\mathrm{b}} = 0.2$ M occurs at $\sigma = -0.0605$ e/nm$^{2}$, as shown in Figure \ref{fig:panel04}. However, the amount of counterions accumulated in the condensed layer is insufficient to overcharge the double layer. Figure \ref{fig:psi_sigma04} shows that the effective surface potential remains negative as the original surface potential. Outside the condensed layer, positive counterions are enriched compared to the bulk (see Figure \ref{fig:conc_sigmac04}), a characteristic of the double layer next to an effective negative surface charge. Furthermore, charge inversion does not occur even with the increase of the surface charge density. In Figure \ref{fig:panel04}, although the condensed layer becomes thicker as $\sigma$ increases, the sign of $\psi_{\mathrm{eff}}$ remains unchanged. These behaviors at two different salt concentrations indicate that ion condensation is necessary for charge inversion but not sufficient. It is also worth noting that our theory predicts ion condensation and charge inversion only for multivalent electrolyte solutions. For monovalent salt systems, the strength of ion correlation is not sufficient to induce ion condensation.\par

\begin{figure}
\centering
 \includegraphics[width=0.6\columnwidth]{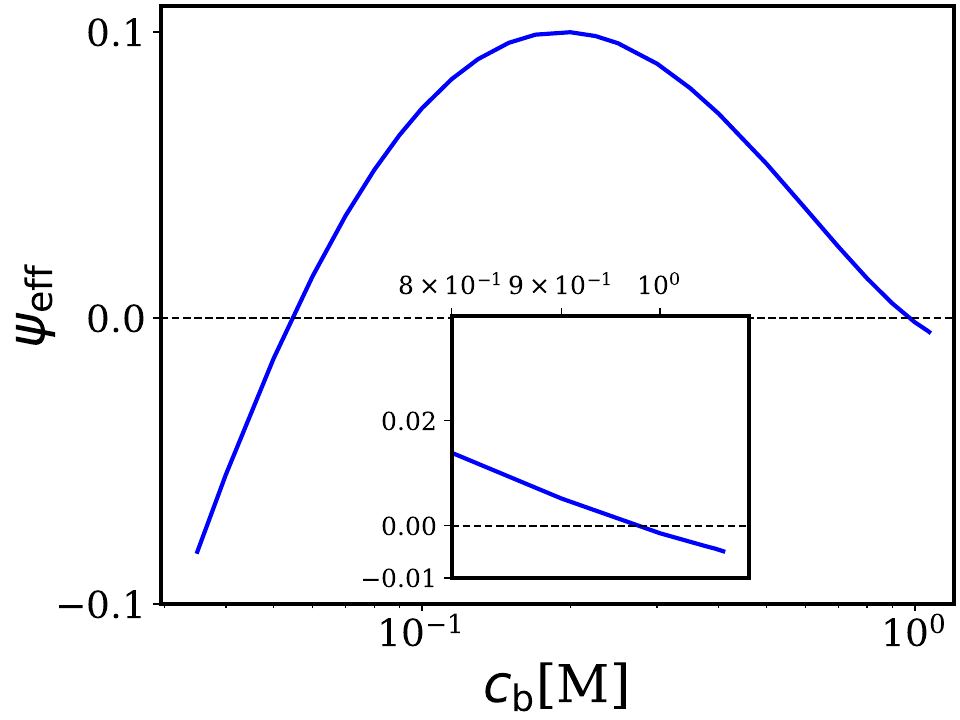}
      \caption{Non-monotonic dependence of effective surface potential ($\psi_{\mathrm{eff}}$) on $c_{\mathrm{b}}$ for $\sigma = -0.1$ e/nm$^{2}$. The inset enlarges the high concentration regime highlighting a second change in the sign of $\psi_{\mathrm{eff}}$. Results here are in qualitative agreement with experiments (Figure 2 of Mart\'in-Molina et al. \cite{Martin-Molina2008ChargeSimulations} and Figure 2 of Van Der Heyden et al.\cite{VanDerHeyden2006ChargeCurrents}) and simulations (Figure 2 of Mart\'in-Molina et al. \cite{Martin-Molina2008ChargeSimulations} and Figure 4 of Hsiao and Luijten \cite{Hsiao2006Salt-inducedPolyelectrolytes})}
            \label{fig:zeta_conc}
\end{figure} 

To get a complete understanding of the relationship between ion condensation and charge inversion, the effective surface potential $\psi_{\mathrm{eff}}$ is plotted for a wide range of bulk salt concentrations. It should be noted that ion condensation occurs in the entire range of $c_{\mathrm{b}}$ for the chosen value of $\sigma$. As shown in Figure \ref{fig:zeta_conc}, $\psi_{\mathrm{eff}}$ has a non-monotonic dependence on $c_{\mathrm{b}}$. At low salt concentrations ($c_{\mathrm{b}}<0.05$ M), the effective surface charge is not inverted in spite of ion condensation. At intermediate salt concentrations, charge inversion occurs. As salt concentration further increases ($c_{\mathrm{b}}>0.97$ M), charge inversion disappears, and $\psi_{\mathrm{eff}}$ regains its original sign. This non-monotonic dependence on salt concentration predicted by our theory is in good agreement with the experimental and simulation results. In experiments, an extremum point was observed in the streaming current measurements inside a nanochannel filled with multivalent electrolyte solutions \cite{VanDerHeyden2006ChargeCurrents}. On the simulation side, electrophoretic measurements of a single polyelectrolyte chain conducted by Hsiao et al.\cite{Hsiao2006Salt-inducedPolyelectrolytes} show a similar non-monotonic trend and disappearance of the inversion of mobility at high concentrations of multivalent ions. \par
\begin{figure}
\captionsetup[subfigure]{labelformat=empty}
    \begin{subfigure}{0.495\columnwidth}
    \includegraphics[width=\columnwidth]{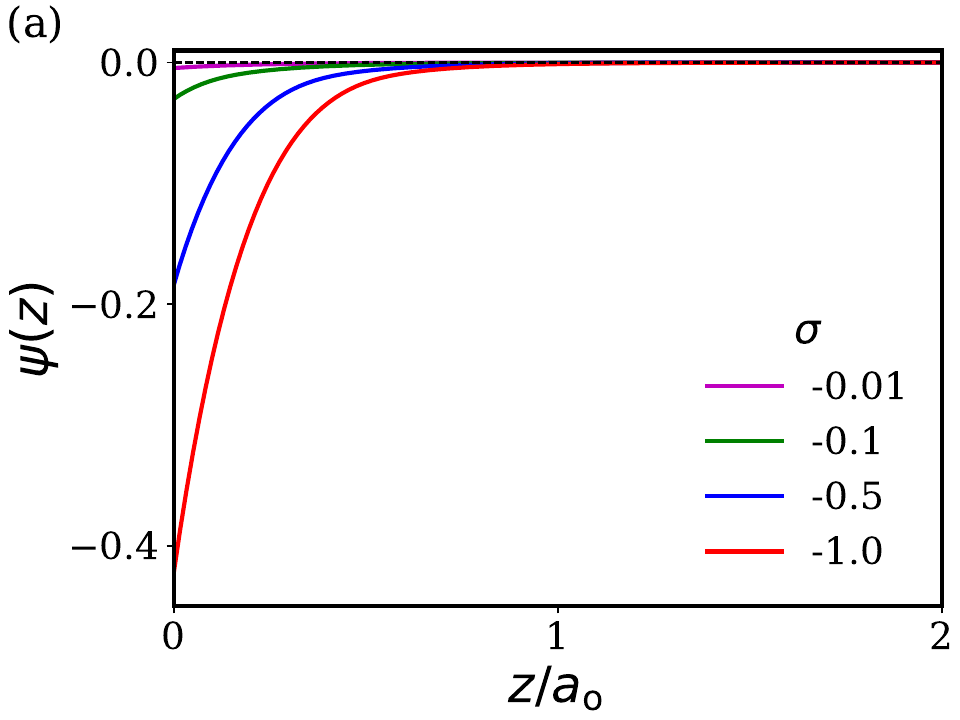}
    \caption{}
    \label{fig:psi_sigma12}
    \end{subfigure}  
    \hfill
    \begin{subfigure}{0.495\columnwidth}
        \includegraphics[width=\columnwidth]{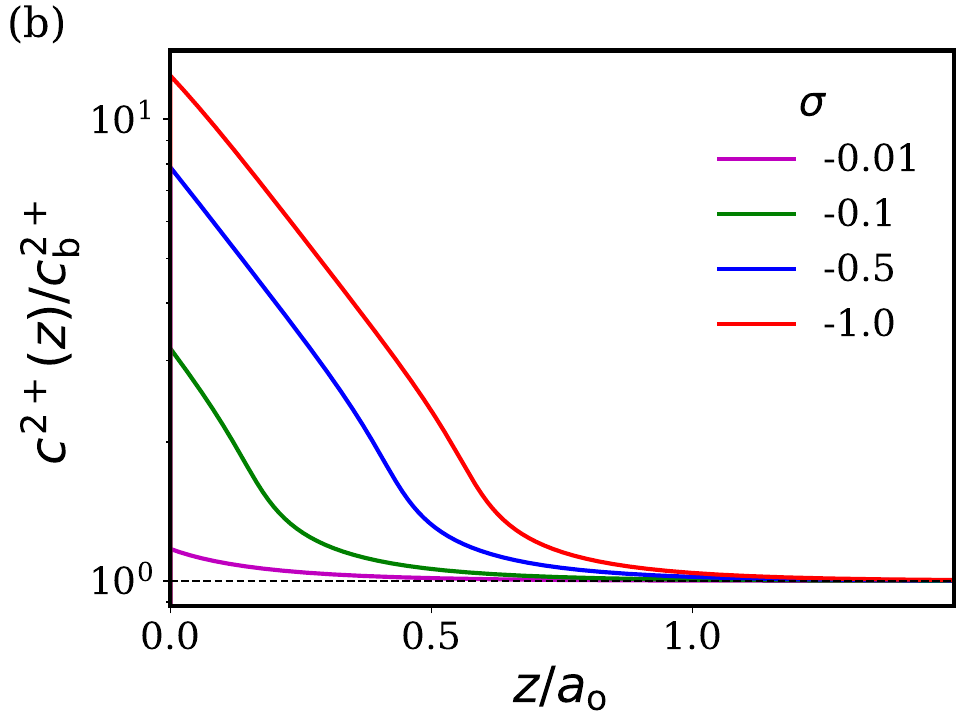}
    \caption{}
        \label{fig:conc_sigmac12}
    \end{subfigure} 
\caption{Double layer structure after the condensed layer dissolves into the diffuse layer at very high salt concentrations. a) Electrostatic potential and b) counterion distribution for $c_{\mathrm{b}}$ = 1.2 M.}
\label{fig:sec_panel}
\end{figure}
The non-monotonic behavior of the effective surface potential can be understood as a consequence of the competition between correlation and the translational entropy of ions. At low salt concentrations ($c_{\mathrm{b}}<0.05$ M), the condensation of ions near the surface leads to a significant entropy loss, which could not be compensated by the energy gain due to ion correlation. Hence, the amount of condensed ions is limited and insufficient to overcharge the surface. As $c_{\mathrm{b}}$ increases, entropy loss reduces for condensing ions, enhancing the migration of ions from bulk to the condensed layer.  $\psi_{\mathrm{eff}}$ increases and crosses zero at $c_{\mathrm{b}} = 0.0545$ M. On the further increase of $c_{\mathrm{b}}$, ion correlation in the bulk region becomes significant, and the driving force for the ions to condense to the surface to lower the self-energy reduces. This strong correlation in bulk results in a maximum in the inverted $\psi_{\mathrm{eff}}$ at $c_{\mathrm{b}}$ = 0.2 M, followed by a decrease in the strength of charge inversion. Eventually, the same effect makes $\psi_{\mathrm{eff}}$ negative again after $c_{\mathrm{b}} = 0.97$ M. Our theory can successfully capture this non-monotonic trend of $\psi_{\mathrm{eff}}$ because it systematically includes the correlation and translational entropy in a unified framework. In addition, the bulk and interface correlations are simultaneously modeled, which is largely missing in existing theories\cite{Bazant2011DoubleCrowding,Perel1999ScreeningInversion,Lau2008FluctuationSolution}. It is important to note that non-monotonic streaming current profiles have been recently reproduced by Gillespie et al. using density functional theory \cite{Gillespie2011EfficientlyTheory}; however, here, we provide the first explanation of the underlying physics of this phenomenon.\par  

Another subtle but important thing to highlight is the dissolving of the condensed layer at very high salt concentrations ($c_{\mathrm{b}}>1.1$ M). In this regime, the bulk salt concentration is so high that there is no significant difference in the strength of ion correlation between bulk and surface. The condensed layer hence dissolves into the diffuse layer. As shown in Figure \ref{fig:psi_sigma12} and \ref{fig:conc_sigmac12}, the electrostatic potential and ion concentration profiles again start behaving in a fashion similar to a normal double layer. Increasing the value of $\sigma$ leads to a monotonic decrease of electrostatic potential and an increase of counterion concentration near the surface. However, unlike double layers in weak coupling conditions, screening lengths are very small in this region due to the high salt concentration. This merging of the condensed layer with the diffuse layer is analogous to the existence of a super-critical phase in the vapor-liquid coexistence. The ability of our theory to capture the dissolution of the condensed layer will play an important role in explaining the behavior of charge inversion in electrolyte mixtures, as will be discussed in the next section.\par

\subsection{Electrolyte mixtures}

Up to now, we have focused on solutions with a single electrolyte. Here, we investigate electrolyte mixtures, which are very important because they widely exist in biological systems as well as geological and environmental systems such as seawater. We take the mixture of 3:1 and 1:1 electrolytes in contact with a negatively charged plate as an example. Figure 6 shows the effect of adding monovalent salt to a trivalent salt solution. The effective surface potential $\psi_\mathrm{eff}$ is plotted as a function of the bulk monovalent salt concentration c$_{\mathrm{b}}^\mathrm{mono}$ for two fixed trivalent salt concentrations (c$_{\mathrm{b}}^\mathrm{tri}=0.005$ M and $0.03$ M). For 3:1 electrolyte solution in the absence of 1:1 salt,  c$_{\mathrm{b}}^\mathrm{tri}=0.005$ M corresponds to the case of condensation without charge inversion, whereas c$_{\mathrm{b}}^\mathrm{tri}=0.03$ M corresponds to the case of condensation with charge inversion. The curves for both of these two c$_\mathrm{b}^\mathrm{tri}$ values show that, with the initial addition of monovalent salt, $\psi_{\mathrm{eff}}$ decreases as c$_{\mathrm{b}}^\mathrm{mono}$ increases, in agreement with experiments\cite{VanDerHeyden2006ChargeCurrents} and simulations\cite{Lenz2008SimulationOvercharging,Quesada-Perez2005SimulationIons}. For c$_{\mathrm{b}}$ = 0.03 M, this decrease of $\psi_{\mathrm{eff}}$ indicates a weakening of charge inversion, which ultimately disappears after c$_{\mathrm{b}}^\mathrm{mono}> 0.013$ M. With the further increase of c$_{\mathrm{b}}^\mathrm{mono}$, the difference between the condensed layer and the diffuse layer becomes insignificant as discussed in the last subsection. \begin{figure}
\centering
\includegraphics[width=0.6\columnwidth]{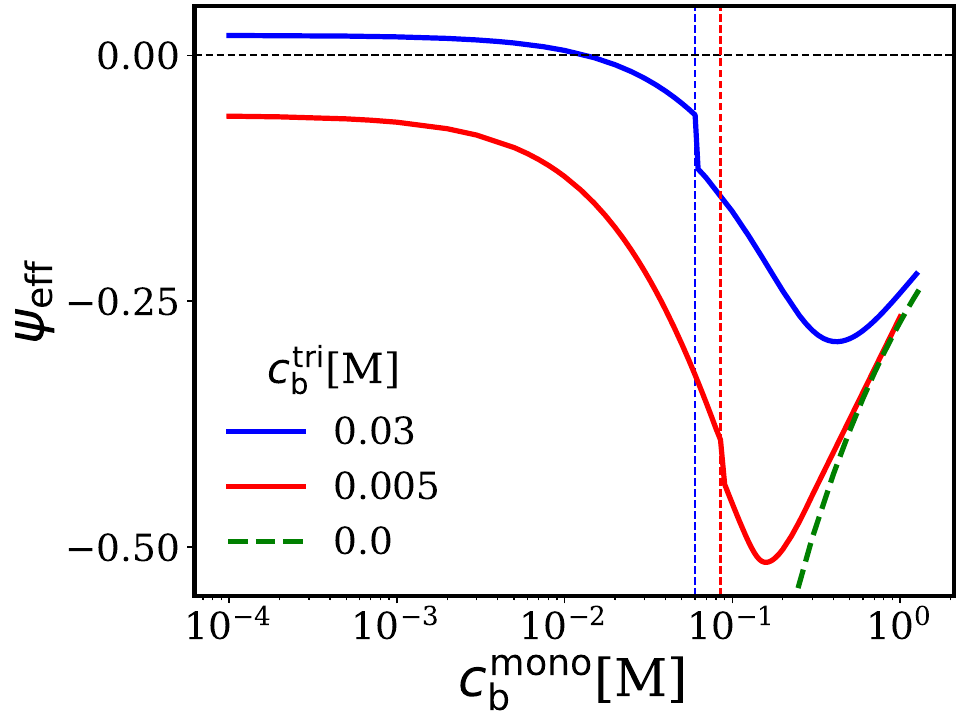}
\caption{Effective surface potential $\psi_{\mathrm{eff}}$ as a function of the concentration of the added monovalent salt, $c_{\mathrm{b}}^\mathrm{mono}$ for fixed values of trivalent salt concentration $c_{\mathrm{b}}^{\mathrm{tri}}$. The vertical dotted lines denote the $c_{\mathrm{b}}^\mathrm{mono}$ at which dissolution of the condensed layer occurs. $\sigma$ is chosen to be -0.1 e/nm$^2$. The Born radius is $a_{\pm}=$ 4 \AA\ . The radius for excluded volumes of ions and solvent is set to be $a_{0}=5$ \AA\. Results here are in qualitative agreement with experiments (Figure 4 of Van Der Heyden et al. \cite{VanDerHeyden2006ChargeCurrents}).}
    \label{fig:zeta_3111} 
\end{figure}   There is a slight drop of $\psi_{\mathrm{eff}}$ in Figure 6 (denoted by vertical dotted lines), which corresponds to the dissolution of the condensed layer.\par

At high concentrations of c$_{\mathrm{b}}^\mathrm{mono}$, $\psi_{\mathrm{eff}}$ reaches a minimum which is followed by an increase. This non-monotonic decrease in $\psi_\mathrm{eff}$ has also been observed by Van Der Heyden et al.\cite{VanDerHeyden2006ChargeCurrents} in streaming current experiments when they added K$^{+}$ salt to a  Co$^\mathrm{3+}$ salt solution. The original strong-coupling (SC) theory by Shklovskii and co-workers predicted that the charge inversion is enhanced by adding monovalent salt, which contradicts these experimental results. Van der Heyden et al. modified SC theory and captured the initial decrease in inversion; however, the non-monotonic trend could still not be reproduced. Our theory provides the first qualitative agreement with experimental results on this non-monotonic behavior in the entire range of monovalent salt concentration.\par 

The initial decrease in $\psi_{\mathrm{eff}}$ implies that with the addition of monovalent salt, the amount of +3 ion condensed in the boundary layer reduces. This can be explained by the competition between the correlation effect and the translational entropy of ions. With the increase of c$_{\mathrm{b}}^\mathrm{mono}$, the ionic correlation in bulk is enhanced. This lowers the driving force for ion condensation because the gain in ionic correlations for trivalent ions when they come to the surface from bulk is not large enough to compensate for the entropy loss. At high c$_{\mathrm{b}}^\mathrm{mono}$, the condensed layer is dissolved into the diffuse layer. In the regime before the minimum, bulk correlation due to the addition of monovalent ions still drives 3+ ions away from the surface to bulk, increasing the magnitude of the effective surface potential. After the minimum, the monovalent salt concentration is so high that the double layer behavior is dominated by monovalent ions, which screen the surface charge. This can be clearly seen in the curve of c$_{\mathrm{b}}^\mathrm{tri}=0.0$ M in Figure \ref{fig:zeta_3111}, where the curves for three c$_{\mathrm{b}}^\mathrm{tri}$ values converge after c$_{\mathrm{b}}^\mathrm{mono}= 1.0$ M. The dominance of monovalent ions at high c$_{\mathrm{b}}^\mathrm{mono}$ was also found in the streaming current experiments \cite{VanDerHeyden2006ChargeCurrents}. Due to this strong screening effect, the surface potential gradually increases and approaches zero. \par 

\section{Conclusions}

We have modified the Gaussian renormalized fluctuation theory by including the excluded volume effect to study ion condensation and charge inversion for multivalent electrolyte solutions in contact with a charged plate. Our results show a surface charge induced formation of a 3D condensed layer of ions near the surface. We have developed a boundary layer approach to capture the giant difference in the ion correlation between the condensed layer near the surface and the diffuse layer outside. In pure multivalent salt solutions, if the amount of condensed ion is large enough, charge inversion occurs. However, we found that ion condensation is necessary but not sufficient for charge inversion. In agreement with experiments\cite{VanDerHeyden2006ChargeCurrents,Martin-Molina2008ChargeSimulations} and simulations\cite{Hsiao2006Salt-inducedPolyelectrolytes}, the effective surface potential exhibits a non-monotonic dependence on the salt concentration. Charge inversion is predicted to only occur in the intermediate salt concentration regime. This non-monotonic behavior of charge inversion, followed by its eventual disappearance, is explained by the competition between the correlation effect and translational entropy of ions. This competition can be accurately modeled by our theory because it systematically includes the inhomogeneous correlations, finite charge spread on the ion, and the translational entropy of solvent through the incompressibility constraint. Our theory provides the first framework to capture both the onset of charge inversion and the double layer structure responsible for the non-monotonic dependence of effective surface potential, hence revealing the underlying physics behind the observations.\par

Furthermore, the effective electrostatic potential for electrolyte mixtures is also investigated. Our calculations show that the initial addition of monovalent salt to a multivalent salt solution results in decay in charge inversion, finally restoring the original sign of effective surface potential. Continued increase of monovalent counterions leads to a minimum in the effective surface potential, which then increases and ultimately approaches zero at very high salt concentrations. Our theory provides the first qualitative agreement with experimental results\cite{VanDerHeyden2006ChargeCurrents} on this non-monotonic behavior of effective surface potential in multivalent and monovalent electrolyte mixtures.\par 

Although the correlation function is solved using WKB approximation at this stage, the remarkable agreement of our calculations with experimental and simulation indicates that our theory has the essential ingredients to capture phenomena originated from ion correlation. The boundary layer model, together with spatially varying self energy, allows us to simultaneously model the correlations at the surface and bulk, an important factor that is largely missing in all the previous theories \cite{Perel1999ScreeningInversion, Bazant2011DoubleCrowding, Lau2008FluctuationSolution}. Here, we have applied our theory to a classic example of electrolyte solutions near a charged surface to address the long-standing puzzle of charge inversion. Both the theory and conclusions of the current work can be generalized to even more complicated systems with correlated/fluctuating charge effects, such as colloidal dispersions, electrokinetic flows, and charged macromolecules. Equation \ref{eq:conc} essentially gives an expression of chemical potential for ions in systems where ion correlations and excluded volume of molecules are expected to play a significant role. This expression for chemical potential can be used in the diffusion term of the Poisson-Nernst-Planck equations to study dynamic phenomena like charging of electrical double layers\cite{bazantchargedynamics2004,guptastonecharging2020}, diffusiophoresis, and diffusioosmosis\cite{guptastonediff2021,suinelectrokinetics2022}. The problem of electrostatic correlation induced like-charge attraction is crucial to understanding the stability of charged colloids\cite{Kumar2017InteractionsIons}; from our equations, the electrostatic forces can be easily calculated for two approaching surfaces. Also, since the equations presented here are derived using a field-theoretic approach, they can be straightforwardly incorporated into self-consistent field theory for polyelectrolytes to understand their structure and interfacial behaviors in response to multivalent salts\cite{Sing2014ElectrostaticMorphology}. Finally, it is important to note that the ion correlations are calculated at the Gaussian level in this work, so the theory will be insufficient to describe systems where higher-order correlations are dominant. Examples include electrolyte solutions at low temperatures \cite{Levin0k_2004} or ionic liquids\cite{gebbieunderscreening2013,guangkornyshevcluster2019} where ion clusters play a critical role.\par

\begin{acknowledgement}

R. W. acknowledges the support from the University of California, Berkeley. This research used
the computational resource provided by the Kenneth S. Pitzer center for theoretical chemistry at UC Berkeley.

\end{acknowledgement}

\bibliography{ci_references.bib}

\end{document}